\documentclass{epl}

\title{
  Dominant particle-hole contributions to the phonon dynamics in the spinless
  one-dimensional Holstein model
}
\shorttitle{Phonon dynamics in the Holstein model}

\author{S. Sykora\inst{1}, A. H\"{u}bsch\inst{2}, and K. W. Becker\inst{1}}

 \institute{
  \inst{1}
  Institut f\"{u}r Theoretische Physik, Technische Universit\"{a}t Dresden,
  01062 Dresden, Germany \\
  \inst{2}
  Max-Planck-Institut f\"{u}r Physik komplexer Systeme, N\"{o}thnitzer
  Stra{\ss}e 38, 01187 Dresden, Germany
}


\begin{document}

\maketitle

\begin{abstract}
In the spinless Holstein model at half-filling the coupling of electrons to
phonons is responsible for a phase transition from a metallic state at small
coupling  to a Peierls distorted insulated state when the electron-phonon
coupling exceeds a  critical value. For the adiabatic case of small phonon
frequencies, the transition is accompanied by a phonon softening at the
Brillouin zone boundary whereas a hardening of the phonon mode occurs in the
anti-adiabatic case. The phonon dynamics studied in this letter do not only
reveal the expected renormalization of the phonon modes but also show
remarkable additional contributions due to electronic particle-hole
excitations. 
\end{abstract}

%%%%%%%%%%%%%%%%%%%%%%%%%%%%%%%%%%%%%%%%%%%%%%%%%%%%%%%%%%%%%%%%%%%%%%%%%%%%%%%
\section{Introduction}

The electron-phonon (EP) interaction leads in many low-dimensional materials
like MX chains, conjugated polymers, or organic transfer complexes 
\cite{mat,mat2} to structural distortions. Thus, Peierls transition and
charge-density wave instability have been observed in such systems. Many
interesting questions arise not only with respect to the associated metal to
insulator transition but also concerning the signatures of the single-particle
excitations in the different phases. The interest in models of
electrons interacting with phonons has been renewed by intriguing findings
which implicate an important role of the EP coupling in a wide range of
materials with strong electronic correlations; high-temperature
superconductors, manganites, or \chem{C_{60}} based compounds are well-known
examples \cite{corr_mat}.

The one-dimensional Holstein model of spinless fermions (HM), 
\begin{eqnarray}
  \label{G1}
  {\cal H} &=&
  - t \sum_{\langle i,j\rangle} ( c_{i}^\dagger c_{j} + \mathrm{h.c.} )
  + \omega_0 \sum_i  \; b_i^\dagger b_i 
  + g \sum_i \; (b_i^\dagger + b_i)n_i,
\end{eqnarray}
is the simplest realization of a strongly coupled EP system and considers the
local interaction $g$ between the electron density 
$n_{i} = c^{\dagger}_{i}c_{i}$ and dispersion-less phonons with frequency
$\omega_0$. The $c^{\dagger}_{i}$ ($b^{\dagger}_{i}$) are fermionic (bosonic)
creation operators of electrons (phonons), and $\langle i,j\rangle$ denotes the
summation over all neighboring lattice sites $i$ and $j$. With increasing $g$, 
the half-filled HM undergoes a quantum phase transition from a metallic
state to a dimerized Peierls phase. Because the HM is not exactly solvable a
number of analytical and numerical methods have been applied (see references
in \cite{Sykora_2006}) to study the phase transition. In particular, the
properties of the insulating phase have been shown to be sensitive to the
relation between band-width and EP coupling \cite{Fehske_2000}: a band
insulator is found in the adiabatic case $\omega_{0}/t \ll 1$ whereas a
polaronic superlattice occurs in the anti-adiabatic limit 
$\omega_{0}/t \gg 1$. 

Recently, we applied the projector-based renormalization method (PRM)
\cite{Becker_2002} to the HM at half-filling where both the metallic and the
insulating case has been studied \cite{Sykora_2005,Sykora_2006}. Here, we
extend our work \cite{Sykora_2006} to discuss the phonon as well as the
electronic one-particle excitation spectrum in more detail. Most of the work
is restricted to the adiabatic limit $\omega_{0}/t \ll 1$ where a phonon
softening is found if the EP coupling approaches the critical value $g_{c}$ of
the transition. In contrast, a phonon stiffening occurs for $\omega_{0}/t \gg
1$. The most remarkable findings of our present work are the large
contributions to the phonon spectrum which are caused by the coupling to
electronic particle-hole excitations. Note that presently in the
anti-adiabatic limit the insulating phase can not be studied because no stable
solutions are found for $t \ll g$.

%%%%%%%%%%%%%%%%%%%%%%%%%%%%%%%%%%%%%%%%%%%%%%%%%%%%%%%%%%%%%%%%%%%%%%%%%%%%%%%
\section{Theoretical approach}
We use the recently derived uniform description of metallic and
insulating phases of the half-filled HM \cite{Sykora_2006}. This approach
employs the PRM \cite{Becker_2002} where an effective Hamiltonian 
$\tilde{\mathcal{H}} = \lim_{\lambda\rightarrow 0} \mathcal{H}_{\lambda}$ is
obtained by a sequence of unitary transformations,
$
  \mathcal{H}_{(\lambda-\Delta\lambda)} =
  e^{X_{\lambda,\Delta\lambda}} \, \mathcal{H}_{\lambda} \,
  e^{-X_{\lambda,\Delta\lambda}}
$,
by which transitions between eigenstates of the unperturbed part
$\mathcal{H}_{0}$ of the Hamiltonian are eliminated in steps. The respective
transition energies are used as renormalization parameter $\lambda$.  
$X_{\lambda,\Delta\lambda}$ has to be adjusted so that
$\mathcal{H}_{(\lambda-\Delta\lambda)}$ only contains excitations with
energies smaller or equal $(\lambda-\Delta\lambda)$. In this way, an
effectively free model, 
\begin{eqnarray}
\label{G2}
  \tilde{ \mathcal{H}} & =&  \sum_{k>0,\alpha} \tilde{\varepsilon}_{\alpha,k}
  c_{\alpha,k}^{\dag} c_{\alpha,k} +
  \sum_{k>0}
  \tilde{\Delta}_{k}^{\mathrm{c}}
  \left(
    c_{0,k}^{\dag} c_{1,k} + \mathrm{h.c.}
  \right) +
  \sum_{q>0, \gamma}
  \tilde{\omega}_{\gamma,q} b_{\gamma,q}^{\dag} b_{\gamma,q} \\
  && + \,
  \sqrt{N} \tilde{\Delta}^{b} \left( b_{1,\pi}^{\dag} + b_{1,\pi} \right) +
  \tilde{E}\, , \nonumber
\end{eqnarray}
was found in Ref.~\cite{Sykora_2006} which is used here in order to calculate
the phonon spectral function,
\begin{eqnarray}
  \label{G3}
  B(q,\omega) &=&
  \frac{1}{2\pi\omega} \int_{-\infty}^{\infty}
  \left\langle [\phi_{q} (t), \; \phi_{q}^\dagger] \right\rangle \;
  e^{i\omega t} \, dt
\end{eqnarray}
and to consider the two electronic one-particle spectral functions
\begin{eqnarray}
  \label{G4}
  A_{k}^{+}(\omega) \,=\, 
  \frac{1}{2\pi} \int_{-\infty}^{\infty}
  \left\langle c_{k} (t) \; c_{k}^\dagger \right\rangle \;
  e^{i\omega t} \,dt\,,  
  \qquad
  A_{k}^{-}(\omega) \,=\, 
  \frac{1}{2\pi} \int_{-\infty}^{\infty}
  \left\langle c_{k}^\dagger \; c_{k}(t) \right\rangle \; 
  e^{i\omega t} \, dt \,. 
\end{eqnarray}
In Eq.~(\ref{G2}) a reduced Brillouin zone has been introduced in order to
allow a dimerization of the system, and both the fermionic and bosonic
one-particle operators as well as the model parameters have additional band
indices, $\alpha, \beta=0,1$. In Eq. (\ref{G3}), $\phi_q = b_q +
b_{-q}^\dagger$ is proportional to the Fourier transformed lattice
displacement. $A_{k}^{+}(\omega)$ describes the creation of an electron with
momentum $k$ at time zero and its annihilation at time $t$ whereas in
$A_{k}^{-}(\omega)$ first an electron is annihilated. As is well-known, 
$A_{k}^{-}(\omega)$ [$A_{k}^{+}(\omega)$] can be measured by [inverse]
photoemission. 

To evaluate Eqs. (\ref{G3}) and (\ref{G4}) within the PRM approach we use that
expectation values are invariant with respect to an unitary transformation
under the trace. Thus, $B(q,\omega)$, $A_{k}^{+}(\omega)$, and
$A_{k}^{-}(\omega)$ can easily be computed if the phononic and electronic
one-particle operators are transformed in the same way as the Hamiltonian (see
Ref. \cite{Sykora_2005} for more details). For instance, by taking the fully
transformed phonon operator
$
  \tilde{b}_{q}^{\dagger} = 
  \tilde{\varphi}_{q} b_{q}^{\dagger} +
  \tilde{\eta}_{q} b_{-q} + 
  \frac{1}{\sqrt{N}}\sum_k \tilde{\psi}_{k,q} c_{k+q}^{\dagger} c_{k}
$
we obtain for the phonon spectral function in the metallic region
\begin{eqnarray}
  \label{G5}
  B(q, \omega) &=& 
  \frac{|\tilde{\varphi}_{q}|^{2}}{\tilde{\omega}_{q} }
  \delta(\omega - \tilde{\omega}_{q}) + 
  \frac{|\tilde{\eta}_{q}|^{2}}{\tilde{\omega}_{-q}}
  \delta(\omega + \tilde{\omega}_{-q}) \\
  && + \,
  \frac{1}{N} \sum_{k} |\tilde{\psi}_{k,q}|^{2}
  \frac{
    f(\tilde{\varepsilon}_{k}) - f(\tilde{\varepsilon}_{k+q}) 
  }{
    \tilde{\varepsilon}_{k+q} - \tilde{\varepsilon}_{k}
  }
  \delta(\tilde{\varepsilon}_{k+q} - \tilde{\varepsilon}_{k} - \omega)
  \nonumber
\end{eqnarray}
where terms with two bosonic creation or annihilation operators have been
neglected. The coefficients $\tilde{\varphi}_{q}$, $\tilde{\eta}_{q}$, and 
$\tilde{\psi}_{k,q}$ can be determined within the PRM approach. Eqs. 
(\ref{G3}) and (\ref{G4}) fulfill sum rules, 
$\int_{-\infty}^\infty \, d\omega \, \omega \, B(q,\omega) = 1$ and
$\int_{-\infty}^{\infty} d\omega \, [A_{k}^{+}(\omega)+A_{k}^{-}(\omega)] =1$,
which are not affected by the PRM approach. All actual calculations are
performed as described in Ref. \cite{Sykora_2006}, and the half-filled HM is
only considered in the one-dimensional case.

%%%%%%%%%%%%%%%%%%%%%%%%%%%%%%%%%%%%%%%%%%%%%%%%%%%%%%%%%%%%%%%%%%%%%%%%%%%%%%%
\section{Adiabatic Limit}

\begin{figure}[t]
  \onefigure{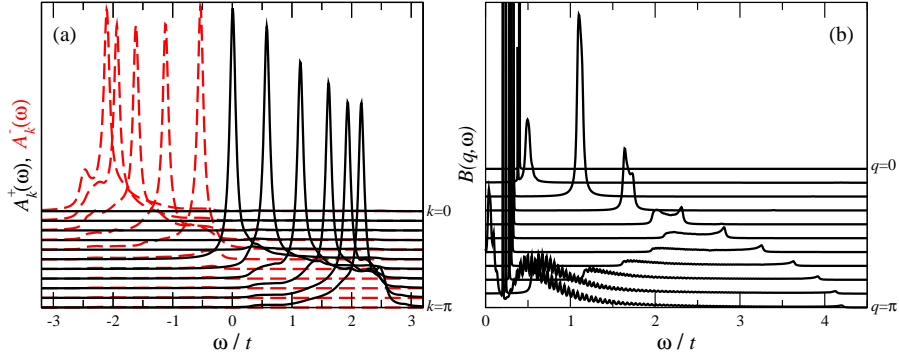}
  \caption{
    (Color online) Panel (a) shows the electronic one-particle spectral
    functions $A_{k}^{+}(\omega)$ (solid lines) and $A_{k}^{-}(\omega)$
    (dashed lines) for the metallic phase in the the adiabatic case,
    $\omega_{0}=0.4\,t$, $g=0.632\,t$ where an one-dimensional HM with 500
    lattice sites has been considered at half-filling. The Fermi energy is
    located at $\omega = 0$. Panel (b) shows the corresponding one phonon
    spectral function $B(q,\omega)$.
  }
  \label{Fig_1}
\end{figure}

In the following we discuss the results for the one-phonon and
one-electron spectral functions, $B(q,\omega)$, $A_{k}^{+}(\omega)$, and
$A_{k}^{-}(\omega)$. We start with the adiabatic limit ($\omega_{0}/t \ll
1$) where the metallic as well as the insulating phase can be described within
the PRM approach. In panel (a) of Fig.~\ref{Fig_1}, $A_{k}^{+}(\omega)$ and
$A_{k}^{-}(\omega)$ are shown for different values of the wave vector $k$ as
functions of $\omega$. The used parameters, $\omega_{0}=0.4\,t$, $g=0.632\,t$,
belong to the metallic region in the adiabatic limit so that no gap occurs at
the Fermi level at $\omega = 0$. Both functions are dominated by the coherent
electronic excitation at $\tilde{\varepsilon}_{k}$
\cite{note_index}. We also find incoherent excitations distributed over an
energy range of about $2 \omega_{0}$. 

As one can see in panel (b) of Fig.~\ref{Fig_1}, the frequency behavior of the 
phonon spectral function $B(q,\omega)$ is also dominated by the coherent
excitation at $\tilde{\omega}_{q}$ \cite{note_index} in this parameter
regime. If the EP coupling $g$ is close to the critical value $g_{c}$
of the metal-insulator transition, a strong softening of the coherent
excitation at $\tilde{\omega}_{q}$ can be observed in the adiabatic limit for
$q$ values approaching the Brillouin zone boundary at $q=\pi$. This phonon
softening is caused by the coupling to electrons and has already been
discussed in Refs. \cite{Sykora_2005,Sykora_2006}. It will turn out that 
a phonon hardening occurs in the anti-adiabatic limit. 

Panel (b) of Fig. \ref{Fig_1} shows that the energies of the incoherent
excitations are distributed over an energy range of the order of the
electronic bandwidth. According to Eq. (\ref{G5}), the energy spreading is
caused by the energy differences of electronic particle-hole excitations 
$(\tilde{\varepsilon}_{k+q} - \tilde{\varepsilon}_{k})$ where
$k$ runs over the whole Brillouin zone. From the numerator in Eq. (\ref{G5}), 
$[f(\tilde{\varepsilon}_{k}) - f(\tilde{\varepsilon}_{k+q})]$, 
one concludes that either $\tilde{\varepsilon}_{k}$ is smaller and
 $\tilde{\varepsilon}_{k+q}$ larger than the Fermi energy or
vice versa. In agreement with Fig.~\ref{Fig_1}, Eq. (\ref{G5}) also shows that 
the  energy range of the incoherent excitations increases with increasing
$q$. Due to the absence of an electronic gap in the metallic
phase, incoherent excitations with small energies are present for $q$-values
close to $q=0$ and $q=\pi$.

\begin{figure}[t]
  \onefigure{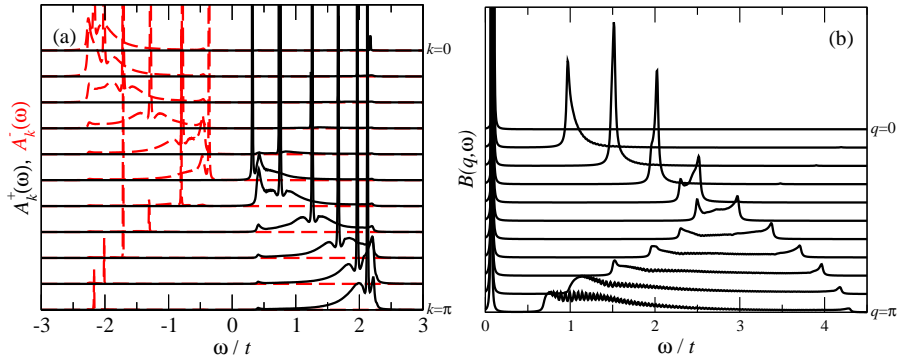}
  \caption{
    (Color online) As Fig. \ref{Fig_1} but for $\omega_{0}=0.1\,t$, $g=0.3\,t$
    (adiabatic case, insulating phase). 
  }
  \label{Fig_2}
\end{figure}

Next, let us consider the ordered insulating phase in the adiabatic limit,
$\omega_{0}/t \ll 1$, where $\omega_{0} = 0.1 \, t$ and $g=0.34\,t$ have been
chosen. The gap of order $\tilde{\Delta}\approx 0.37\,t$ in the electronic
excitation spectrum [compare panel [a] of Fig. \ref{Fig_2}] clearly shows that
the system is in the insulating state. Again the electronic spectral functions
$A_{k}^{+}(\omega)$ and $A_{k}^{-}(\omega)$ are dominated by the coherent
excitation at $\tilde{\varepsilon}_{k}$, and the spreading of the incoherent
excitations is similar to that of the metallic state [compare with panel (a)
of Fig. \ref{Fig_1}]. As one can see in panel (b) of 
Fig. \ref{Fig_2}, the spectral weight of the coherent excitation is for
the phonon spectral function $B(q,\omega)$ again huge in comparison to that of
the incoherent excitations. However, it is important to notice a significant
difference to the metallic state [compare panel (b) of Fig. \ref{Fig_1}]:
No incoherent excitations occur at low energies which corresponds to the gap
in the electronic spectrum.

%%%%%%%%%%%%%%%%%%%%%%%%%%%%%%%%%%%%%%%%%%%%%%%%%%%%%%%%%%%%%%%%%%%%%%%%%%%%%%%
\section{Anti-adiabatic Limit}

\begin{figure}[t]
  \onefigure{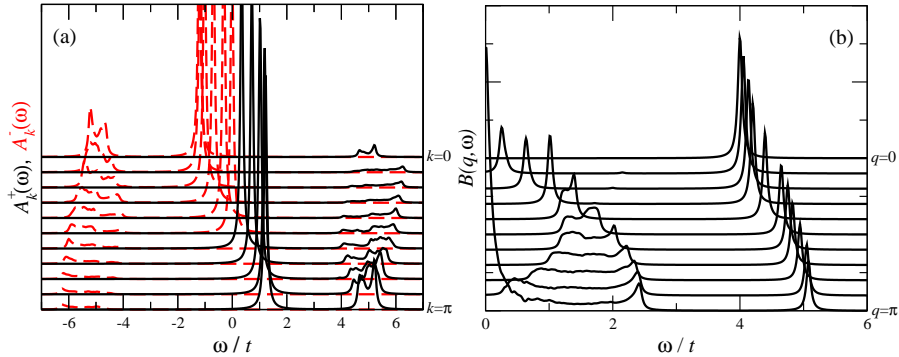}
  \caption{
    (Color online) As Fig. \ref{Fig_1} but for $\omega_{0}=4\,t$, $g=2.82\,t$
    (anti-adiabatic case, metallic phase).
  }
  \label{Fig_3}
\end{figure}

Next we discuss the anti-adiabatic limit where $\omega_{0}/t \gg 1$
holds. Unfortunately, in this limit we are restricted to the metallic case 
$g < g_{c}$, because no stable solution for the insulating phase $g > g_{c}$ 
is found within the PRM approach. The reason is that the absolute
value of the EP coupling $g$ is too large in this case. As one can see in
Ref. \cite{Sykora_2006}, the renormalization equations were derived by
starting from an uncorrelated model by successively eliminating high-energy
excitations which prevents the application of the PRM scheme for extremely
high coupling parameters $g$. The electronic spectral functions 
$A_{k}^{+}(\omega)$ and $A_{k}^{-}(\omega)$ are shown in panel (a) of
Fig. \ref{Fig_3} for the metallic phase, where $\omega_{0}=4\,t$ and
$g=2.82\,t$ have been chosen. The missing gap in the electronic spectrum
reveals the metallic state of the system, and the $k$ dependence of the sharp
coherent excitation peak corresponds to the dispersion
$\tilde{\varepsilon}_{k}$ of the electrons. The coherent excitation is
separated from the incoherent excitations by an energy of the order of the
phonon energy $\omega_{0}$. Here, a renormalized one-electron creation operator
$c_{k}^{\dag}(\lambda)$ has been used which includes phonon operators only in
linear order (see Ref. \cite{Sykora_2005} for details). This coupling
leads to the energetic separation of order $\omega_{0}$. If operator
terms with more than a single phonon operator were taken into account
additional excitations at higher frequencies ($2\omega_{0}$, $3\omega_{0}$,
etc.) would appear. This is confirmed by recent ED results of
\cite{Hohenadler}. 

The metal-insulator transition in the anti-adiabatic limit can be understood 
as the formation of small immobile polarons which are electrons surrounded by
clouds of phonon excitations. In the PRM approach polarons correspond with
the fully renormalized electronic excitations $\tilde{\varepsilon}_{k}$. If
one compares the renormalized electronic band width for the adiabatic
case, which can be read off from the coherent excitation in
panel (a) of Fig. \ref{Fig_1}, with that of Fig. \ref{Fig_3} in the
anti-adiabatic case one observes a strong reduction of the band width. 
This reduction clearly indicates localization tendencies in
the system which might allow the determination of the critical EP coupling
$g_{c}$ of the metal-insulator transition in the anti-adiabatic limit within
our PRM approach \cite{Sykora_unpub}.

Panel (b) of Fig.~\ref{Fig_3} shows the phononic spectral function
$B(q,\omega)$ in the anti-adiabatic limit. It is important 
to notice three significant
differences to the adiabatic case [compare with panel (b) of
Fig. \ref{Fig_1}]: Firstly, the coherent phonon excitation no longer shows a
softening behavior, instead an hardening of the phonon modes is observed from
a value $\omega_{0}$ at $q=0$ to higher energies for all $q>0$. Secondly, the
incoherent contributions from particle-hole excitations have gained
considerable weight as compared to the adiabatic case [see panel (b) of
Fig. \ref{Fig_1}]. The energy range of the incoherent excitations is again of
the order of the electronic band width. Finally, a huge elastic contribution
is found in $B(q,\omega)$ for $q=\pi$.

%%%%%%%%%%%%%%%%%%%%%%%%%%%%%%%%%%%%%%%%%%%%%%%%%%%%%%%%%%%%%%%%%%%%%%%%%%%%%%%
\section{Discussion}

Phononic and electronic spectral functions for the one-dimensional HM at
half-filling were recently evaluated in Ref. \cite{Hohenadler}. 
In this work, the
same parameter values as used here were studied by exact diagonalization
techniques (ED) for system sizes of up to $10$ lattice sites. The results for
the electronic spectral functions $A_{k}^{+}(\omega)$ and $A_{k}^{-}(\omega)$
agree quite well with our analytical results in the whole metallic
regime. In the phonon spectrum two essential differences are found. 
In the ED-approach \cite{Hohenadler} a strong excitation is found at
frequency $\omega = 0$ for wave vector $q=0$ both in the adiabatic and in the
anti-adiabatic case. In contrast, the results in Figs. \ref{Fig_1} and
\ref{Fig_3} show excitations for small $q$ only for finite frequencies
$\omega>0$. This additional zero-frequency peak in Ref. \cite{Hohenadler} is
caused by a different definition of the phonon spectral function. Instead of
the commutator spectral function $B(q,\omega)$, in \cite{Hohenadler} a
phononic correlation function,
$
  \tilde{B}(q,\omega) =
  \frac{1}{2\pi\omega} \int_{-\infty}^{\infty}
  \left\langle \phi_{q} (t) \; \phi_{q}^{\dagger} \right\rangle
  \; e^{i\omega t} dt
$,
is investigated so that an additional zero-frequency contribution from the
center of gravity motion contributes, which is dropped in the commutator
spectral function as defined in Eq. (\ref{G3}).

\begin{figure}[t]
  \onefigure{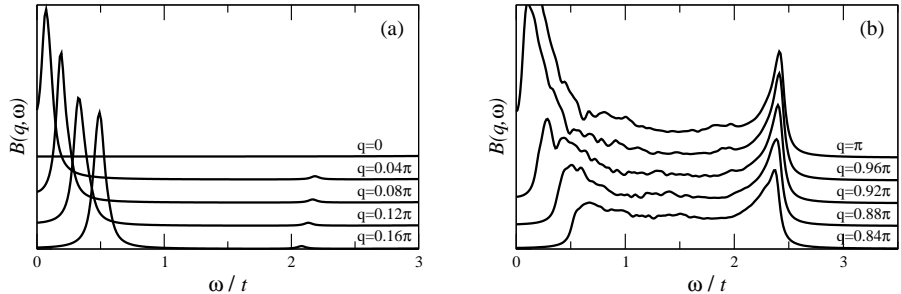}
  \caption{
    Low energy range of the phonon spectral function as shown in
    Fig. \ref{Fig_3} for small $q$ values [panel (a)] and for $q$ values close
    to the Brillouin-zone boundary [panel (b)].
  }
  \label{Fig_4}
\end{figure}

The second difference between our results and those of the ED approach of
Ref. \cite{Hohenadler} is more important: In the phonon
spectrum of \cite{Hohenadler} almost no small frequency excitations are found
in the vicinity of $q=\pi$ whereas the present PRM approach for
$B(q=\pi,\omega)$ in the anti-adiabatic limit gives a strong zero-energy peak
at $q=\pi$. For a more detailed discussion let us consider the phonon spectrum
for different $q$-values in the vicinity of $q=\pi$ [panel (b) of
Fig.~\ref{Fig_4}]. As follows from (\ref{G5}), the incoherent excitations in
$B(q,\omega)$ are caused by the coupling of the phonons to electronic
particle-hole excitations with energies 
$|\tilde{\varepsilon}_{k+q} - \tilde{\varepsilon}_{k}|$ where 
the wave vector $k$ runs over the whole Brillouin zone.
The energies $\tilde{\varepsilon}_{k+q}$ and $\tilde{\varepsilon}_{k}$
have to be below and above the Fermi level with momentum $k_{\mathrm{F}}$. 
Thus, only electronic particle-hole excitations contribute
with either $|k| < k_{\mathrm{F}}$ and $|k+q| > k_{\mathrm{F}}$, or 
$|k|>k_{\mathrm{F}}$ and $|k+q| < k_{\mathrm{F}}$. For $q=\pi$ this condition
is fulfilled for all $k$-values of the Brillouin zone, and incoherent
excitations are found in an energy regime of the order of the electronic band
width. For $q=\pi$ the smallest possible energy difference 
$|\tilde{\varepsilon}_{k+q} - \tilde{\varepsilon}_{k}|$ is the same as
the energy difference of two neighboring $k$ points around $k_\mathrm{F}$
with the smallest possible distance in $k$-space. This distance is of
order $1/N$ and might be not small for a small system size. 
This explains the absence of low-energy excitations in the phonon spectrum of
Ref. \cite{Hohenadler} close to the Brillouin-zone boundary (where only system
sizes of order N=10 have been considered). We could confirm this mechanism
within our PRM approach: If we reduce the system size from 500 to 66 lattice
sites the low-frequency peak in $B(q=\pi,\omega)$ is almost completely
suppressed, and its position is shifted to higher energies.

Let us also discuss the case of small $q$-values $q \ll \pi$, where 
only $k$ values from the sum in (\ref{G5}) can contribute which are located in
a small wave vector region around the Fermi momentum
$k_{\mathrm{F}}$. Assuming a linear momentum dependence for energies close to
Fermi energy, the incoherent electronic excitation energies can be replaced by 
$
  |\tilde{\varepsilon}_{k+q} - \tilde{\varepsilon}_{k}|\approx
  v_{\mathrm{F}}q
$. 
($v_{\mathrm{F}}$ is the Fermi velocity.) For small $q$ the
position of the incoherent excitations is proportional to $q$, as seen
in panel (a) of Fig.~\ref{Fig_4}. On the same time the intensity of the
incoherent peak is approximately $q$ independent which follows again from
(\ref{G5}): The linear $q$ dependence of the denominator is canceled by the
number of contributing $k$ points which is also proportional to $q$ so that an
approximately $q$ independent intensity results. 

Finally, we use the PRM to evaluate the Luttinger liquid parameters $u_\rho$,
$K_\rho$ in the metallic phase at half-filling. It is commonly accepted that
the spinless one-dimensional HM belongs to the Tomonaga-Luttinger
universality class, and at $T=0$ a Luttinger liquid should be realized. Due to
the quite different behavior of the phonon spectral function in the adiabatic
and in the anti-adiabatic regime one expects different results in both
cases. For the velocity $u_\rho$ of the charge excitations, we use a
finite-size scaling of the energy gap~\cite{Schulz,Noack}, 
$
  \Delta _\rho = \frac{1}{2}( E_0^{+1} +  E_0^{-1} -2 E_0) = \frac{\pi}{N}
  \frac{u_\rho}{2}
$,
where $N$ is the lattice size, and $E_0$ denotes the ground-state energy of
the half-filled system. $ E_0^{\pm 1}$ is the ground-state energy of the
system with $\pm 1$ fermions away from half-filling. To compute the effective
coupling constant $K_\rho$ we take the first derivative of the charge
correlation function 
$C(q)$ at $q=0$,
$ 
  \frac{\partial C(q)}{\partial q}|_{q=0} = \frac{K_\rho}{\pi}
$,
where 
$
  C(q)= \frac{1}{N} \sum_{i(\neq)j} 
  e^{iq(R_i-R_j)} \langle \delta n_i \delta n_j \rangle
$.  
Results are given in Table \ref{Tab}. In the anti-adiabatic
regime the parameter $u_\rho$ for the kinetic energy is strongly reduced when
$g$ is increased which corresponds to the band width reduction as discussed
above [compare panel (a) of Fig.~\ref{Fig_3}]. This behavior of $u_\rho$ is
very similar to recently found results \cite{Fehske_2004}. In contrast, the
results for $K_\rho$ represent an unsolved puzzle: Here, $K_\rho$ is always
smaller than $1$, whereas in Ref.~\cite{Fehske_2004} $K_\rho$ was  
smaller than $1$  only in the anti-adiabatic regime but larger than $1$ 
in the adiabatic regime.
 
\begin{table}
  \begin{center}
    \begin{tabular}{c|ccc|ccc} \hline\hline
      & $\phantom{\omega_{0}/t = 0.1}$ & $\omega_{0}/t = 0.1$ 
      & $\phantom{\omega_{0}/t = 0.1}$ & $\phantom{\omega_{0}/t = 0.1}$ 
      & $\omega_{0}/t = 4.0$ & $\phantom{\omega_{0}/t = 0.1}$ \\ \hline
      $g/t$ & 0.078 & 0.141 & 0.2 & 1.0 & 1.8 & 2.82 \\
      $K_{\rho}$ & 0.966 & 0.888 & 0.882 & 0.928 & 0.790 & 0.740 \\
      $u_{\rho} / 8t$ & 0.978 & 0.933 & 0.928 & 0.935 & 0.708 & 0.647 \\
      \hline\hline
    \end{tabular}
  \end{center}
  \caption{
    LL parameters in the adiabatic and in the anti-adiabatic
    regime. To account for  the missing spin degree of freedom, 
    $K_\rho$ has been multiplied by a factor of 2.
  }
  \label{Tab}
\end{table}

%%%%%%%%%%%%%%%%%%%%%%%%%%%%%%%%%%%%%%%%%%%%%%%%%%%%%%%%%%%%%%%%%%%%%%%%%%%%%%%
\section{Summary}

In this paper we have considered the phononic and electronic spectral functions
of the one-dimensional HM at filling. In particular, we have studied the
spectral signatures of the metallic phase both in the adiabatic and the
anti-adiabatic limit, and of the insulating phase in the adiabatic
limit. (The insulating state could not be investigated in the anti-adiabatic
limit because the employed PRM approach breaks down in this case.) We find a
quite strong coupling of the phonon dynamics to electronic particle-hole
excitation which leads to incoherent contributions to the phononic spectral
function. In particular, a dominant incoherent low-energy peak is observed for 
the metallic phase in the anti-adiabatic limit.

\acknowledgments
We would like to acknowledge helpful discussions with H.~Fehske. This work
was supported by the DFG through the research program SFB 463.


\begin{thebibliography}{0}
\bibitem{note_index}
  Here we suppressed the band index $\alpha$ for the one-particle energy
  because it is not needed for metallic solutions. See Ref. \cite{Sykora_2006}
  for details.
\end{thebibliography}
\end{document}